\documentclass[preprint,preprintnumbers,amsmath,amssymb]{revtex4}

\usepackage{graphicx}
\usepackage{dcolumn}
\usepackage{bm}
\usepackage{verbatim}
\usepackage{epsfig}
\usepackage{epstopdf}
\usepackage{amsmath}
\usepackage[caption=false]{subfig}	
\usepackage{dsfont}

\newcommand\beq{\begin{equation}}
\newcommand\eeq{\end{equation}}
\newcommand\bea{\begin{eqnarray}}
\newcommand\eea{\end{eqnarray}}

\newcommand\expval[1]{\langle #1 \rangle}

\usepackage{xcolor}

\def\P{{\bf P}}

\def\id{\mathds{1}}
\def\half{\frac {1} {2}}

\def \MR {\rm MR}
\def\LG{\rm LG}
\def\NSIT{\rm NSIT}
\def\NIM{\rm NIM}
\def\Ind{\rm Ind}
\def\MRps{\rm MRps}

\def\P{{\mathbb E}}

\def\x0{{{\bf x}_0}}

\begin{document}
	
	
	\title{Conditions for Macrorealism for Systems Described by Many-Valued Variables}

	\author{J.J.Halliwell}%
	\email{j.halliwell@imperial.ac.uk}
	\author{C. Mawby}
	\email{c.mawby18@imperial.ac.uk}
	\numberwithin{equation}{section}
	\renewcommand\thesection{\arabic{section}}

	\affiliation{Blackett Laboratory \\ Imperial College \\ London SW7
		2BZ \\ UK }

	
\begin{abstract} 
Macrorealism (MR) is the view that a system evolving in time possesses definite properties independent of past or future measurements and is traditionally tested for systems described at each time by a single dichotomic variable $Q$. A number of necessary and sufficient conditions for macrorealism have been derived for a dichtomic variable
using sets of Leggett-Garg (LG) inequalities, or the stronger no-signaling in time (NSIT) conditions, or a combination thereof. Here, we extend this framework by establishing necessary and sufficient conditions for macrorealism for measurements made at two and three times for systems described by variables taking three or more values at each time. Our results include a generalization of Fine's theorem to many-valued variables
for measurements at three pairs of times 
and we derive the corresponding complete set of LG inequalities. 
We find that LG inequalities and NSIT conditions for many-valued variables do not enjoy the simple hierarchical relationship exhibited by the dichotomic case. This sheds
light on some recent experiments on three-level systems which exhibit a LG inequality violation even though certain NSIT conditions are satisfied.
Under measurements of dichotomic variables using the L\"uders projection rule the three-time LG inequalities cannot be violated beyond the L\"uders bound (which coincides numerically with the Tsirelson bound obeyed by correlators in Bell experiments), but this bound can be violated in LG tests using degeneracy-breaking (von Neumann) measurements.  We identify precisely which MR conditions are violated under these circumstances.
\end{abstract}


	
	\maketitle

	\section{Introduction}

\subsection{Background}

Much current research in quantum theory is devoted to identifying genuinely quantum-mechanical effects that elude any kind of classical explanation. There are at least two reasons why this is of interest. Firstly, to test fundamentals, and address such questions as to whether quantum coherence persists to the macroscopic domain. Secondly, genuinely quantum effects can be exploited as resources which encourage the development of new technologies.

An important example of this endeavour is the Leggett-Garg (LG) framework, which tests a specific notion of classicality called macrorealism (MR) \cite{LeGa,L1}. (See Ref.\cite{ELN} for an extensive review).
This is the view that a system evolving in time possesses definite properties independent of past or future measurements. Most theoretical and experimental investigation of the LG framework concern measurements on a single dichotomic variable $Q$ evolving in time. Non-invasive measurements of $Q$ are made in three experiments, each involving three pairs of times, which permit the determination of a set of temporal correlation functions of the form,
\beq
C_{12} = \langle Q_1 Q_2 \rangle,
\label{corr}
\eeq
where $Q_i$ denotes the value of $Q$ at time $t_i$, and the average is over a large number of experimental runs of the product of the sequentially measured values of $Q_1$ and $ Q_2$.
It is argued that for a macrorealistic theory, the correlations functions arising from measurements at three pairs of times must obey the LG inequalities:
\bea
1 + C_{12} + C_{23} + C_{13} & \ge & 0,
\label{LG1}
\\
1 - C_{12} - C_{23} + C_{13} & \ge & 0,
\label{LG2}
\\
1 + C_{12} - C_{23} - C_{13} & \ge & 0,
\\
1 - C_{12} + C_{23} - C_{13} & \ge & 0.
\label{LG4}
\eea
These inequalities follow from breaking MR down into three separate assumptions:
the variable $Q$ takes a definite value at each time (macrorealism per se, MRps); that they can be measured without disturbing the future evolution of the system (non-invasive measurability, NIM); and that future measurements cannot affect the past (induction, Ind). In brief,
\beq
\MR = \NIM \wedge \MRps \wedge \Ind.
\label{MRdef}
\eeq
(Here $\wedge$ denotes logical conjunction).
These assumptions imply that the values of $Q$ at the three times $t_1$, $t_2$, $t_3$ possess a joint probability distribution and the LG inequalities readily follow.
Many different experiments on a variety of different systems have been carried out to test the LG inequalities and violations in accordance with the predictions quantum mechanics have been observed. (See for example the review Ref.\cite{ELN}).

\subsection{Necessary and Sufficient Conditions for Macrorealism}

The LG framework was developed by way of analogy to Bell experiments \cite{Bell,CHSH}
(although this analogy is at best partial \cite{MaTi}).  For Bell experiments on an entangled pair of qubits, Fine's famous theorem ensures that the Bell or CHSH inequalities are both necessary and sufficient conditions for the existence of an underlying probability matching the pairwise marginals determined by the measurements \cite{Fine,Bus,SuZa,Pit,GaMer,ZuBr,JHFine,AbBr}.
 This is not the case for the LG inequalities, Eqs.(\ref{LG1})-(\ref{LG4}) which are only necessary conditions for MR.
The shortfall may however be made up by adjoining the four LG inequalities with a set of twelve two-time LG inequalities of the form,
\beq
1 + s_i \langle Q_i \rangle + s_j \langle Q_j \rangle + s_i s_j C_{ij} \ge 0, 
\label{2time}
\eeq
where $s_i = \pm 1 $, $i,j$ take values $1,2,3$ with $ i < j $, and the three averages $\langle Q_i \rangle $ are measured using three additional experiments.
The set of sixteen inequalities consisting of Eqs.(\ref{LG1})-(\ref{LG4}) together with Eq.(\ref{2time}) constitute a set of necessary and sufficient conditions for macrorealism \cite{HalQ,HalLG4}. Almost all experimental tests check only a subset of this set, so test only necessity, but an experimental test of the complete set of inequalities was recently carried out \cite{Maj,MajThe}. This framework, involving sets of necessary and sufficient conditions for MR, was also recently extended to measurements at an arbitrary number of times \cite{HaMa} and to situations involving higher order correlators \cite{HalNIM}  (considered earlier in Ref.\cite{PQS}).
Conditions of this general type have also been derived independently in polytope constructions \cite{poly}.

However, it is important to emphasize that the definition of MR indicated by Eq.(\ref{MRdef}) is open to a number of different interpretations due to the fact that there are a number of different and physically reasonable ways of interpreting both MRps \cite{MaTi} and NIM \cite{HalLG4,KoBr,Cle}. In Ref.\cite{HalLG4}, a number of different notions of MR were identified, corresponding to different ways of implementing NIM. In this classification,
the above characterization of MR using LG inequalities at two and three times, with the $\langle Q_i \rangle$ and $C_{ij}$ measured in six experiments, is referred to as {\it weak} MR.

The other important class of MR conditions
are those involving no-signaling in time (NSIT) conditions, which, at two times entail the determination of a probability $p_{12} (s_1, s_2)$ through two sequential measurements of $Q$ in a single experiment and requiring that
\beq
\sum_{s_1} p_{12} (s_1, s_2) = p_2 (s_2),
\label{NSIT}
\eeq
where $p_2(s_2)$ is the probability for $Q$ at $t_2$ with no earlier measurement at $t_1$ \cite{KoBr}. Analogous conditions at three or more times are readily constructed \cite{Cle}. Such conditions also ensure the existence of an underlying probability, but as argued in Refs.\cite{HalQ,HalLG4} these conditions test a {\it different notion} of macrorealism which is stronger than that characterized purely by LG inequalities. Characterizations of MR which entail only NSIT conditions are therefore referred to in Ref.\cite{HalLG4} as {\it strong} MR. Intermediate possibilities also exist involving combinations of LG inequalities and NSIT conditions. All of these notions of MR are of interest and as we shall see there is some interplay between the two.
(Furthermore, these possibilities are not the only ones -- see for example 
Refs.\cite{WLG,KuPa}).

The above descriptions of weak and strong MR are sufficient background for most of what follows in this paper. However, for convenience a more extensive description summarizing and extending the results of Ref.\cite{HalLG4} is given in Appendix A.

\subsection{This Paper}

Almost all experimental studies of the LG framework concern small systems and the original goal of using the framework to study coherence in macroscopic systems is still on the far horizon. (Although see the recent proposal Ref.\cite{Bose}). However, as the systems studied become progressively larger, it becomes of interest to extend the above frameworks to embrace the new features of larger systems not present, for example, in simple two-state systems, or for systems described by a single dichotomic variable.

The purpose of the present paper is to develop generalizations of the above necessary and sufficients tests of macrorealism for the case of many-valued variables.
We thus imagine doing a set of experiments each of which may involve one, two or three times, and which determine certain averages and correlators of an $N$-valued variable. 
This includes both the case of fine-grained measurements on an $N$-level system, or coarse-grained measurements on systems  with dimension greater than $N$.
Taking first the case of weak MR,
these measurements provide a determination, indirectly, of a set of candidate two-time probabilities of the form $p(n_1, n_2)$,
$p(n_2,n_3)$, $p(n_1,n_3)$ for the three time pairs $(t_1,t_2)$, $(t_2,t_3)$, $(t_1,t_3)$, where $n$ runs from $1$ to $N$ and $N \ge 3$. The question as to whether these candidate two-time probabilities are non-negative leads to the desired generalization of the two-time LG inequalities, Eq.(\ref{2time}).
From there one can ask for the conditions under which these three pairwise probabilities are the marginals of a single three-time probability $p(n_1,n_2,n_3)$. We will thus develop the appropriate generalizations of the three-time LG inequalities and Fine's theorem. We also develop definitions of strong MR by exploring the generalizations of the NSIT conditions.
We will use our analysis of LG inequalities and NSIT conditions to explore some of the novel features of MR conditions for many-valued variables not present in the dichotomic case.

Some of the finer details of the MR conditions derived in the following sections are quite complicated. However, we stress that the underlying strategy is quite simple: in each case we will show how to express measurements on a many-valued variable in terms of measurements on a related set of dichotomic variables, for which the MR conditions are closely related to those given above.

In Section 2, we review commonly-used conditions for MR for measurements of a single dichotomic variable as outlined above. 
The discussion is primarily from a 
quantum-mechanical perspective, from which such conditions are seen to be restrictions on the degree of interference between different histories of the system. This helps in determining the degree to which different MR conditions are independent of each other and also illustrates the link between LG inequalities and NSIT conditions, preparing the way for the many-valued case.

The extension of the standard approach involving LG inequalities for dichotomic variables to many-valued variables is presented in Sections 3 and 4. We first, in Section 3, consider the simpler case of MR conditions at two times and determine the form of the LG inequalities for variables taking $N$ values at each time.
In Section 4 we extend our results to the standard LG situation in which measurements are made at three pairs of times. We prove the generalization of Fine's theorem required for this case, i.e. we determine necessary and sufficient conditions for the existence of an underlying joint probability matching the measured two-time marginals for $N$-valued variables.
The conditions turn out to be a set of LG inequalities which can be written quite simply in the general case.
We note that our results are readily extended to measurements at many times by taking advantage of the generalized Fine ansatz presented in Ref.\cite{HaMa}.

In Section 5 we consider the different and stronger MR conditions involving NSIT conditions, primarily focusing on the two-time case and we discuss the logical relationships between such conditions and the weaker LG inequalities. We find that it is considerably less simple than the dichotomic case and
we use this understanding to analyze the results of some recent experiments in which two-time LG violations are observed even though certain NSIT conditions are satisfied. 

In Section 6, we use the methods developed above to examine the surprising fact that the LG framework sometimes permits violations of the LG inequalities to a degree beyond the so-called 
L\"uders bound (which is $- \frac{1}{2}$ in the right-hand side of Eqs.(\ref{LG1})-(\ref{LG4})),
using particular types of measurements available only for $N \ge 3$ \cite{Dak,deg,EmaExp,PQS,KQP}.
This bound coincides numerically with the Tsirelson bound for the correlators in Bell experiments \cite{Tsi} but for LG tests the L\"uders bound is not in fact the maximum permitted by quantum mechanics.
We show that
the violation naturally separates into a conventional LG violation (respecting the L\"uders bound) together with a violation of a two-time NSIT condition and discuss the consequences of this in terms of MR conditions.
We summarize and conclude in Section 7.

As mentioned, Appendix A contains a more detailed account of the different types of definitions of MR for dichotomic variables alluded to above \cite{MaTi,Cle,HalQ,HalLG4}. 
Some useful technical results from the decoherent histories approach to quantum theory are given in Appendix B.  
In Appendix C we record for future use the full list of LG inequalities for the $N=3$ case at three times, a list too lengthy for the main text.

\section{Conditions for Macrorealism for Dichotomic Variables}


The  LG framework is usually described from a purely macrorealistic point of view in which the experimental situation is a black box, about which little is assumed in terms of the dynamics, initial state etc.  Although we will follow this approach where possible,
in this paper we find that a quantum-mechanical analysis is often most convenient, since the LG inequalities and NSIT conditions that we seek to derive have a simple interpretation in quantum mechanics as restrictions on the size of certain interference terms, as will become clear in what follows (and see also Ref. \cite{HalQ}). 
The resulting understanding may subsequently be re-expressed in purely macrorealistic terms but we do not always spell this out explicitly.


In this section we describe the quantum-mechanical description of MR conditions for a single dichotomic variable at two and three times. Section 2(A) is primarily a summary of earlier work \cite{HalQ}, with some additional commentary,
but the details of this work are important for the generalizations considered in later sections.

\subsection{Two-Time LG Inequalities, NSIT Conditions and Interferences}

We imagine a system with Hamiltonian $H$ subject to measurements at either single times in each experiment or pairs of times. 
Measurements of the dichotomic variable $Q$ are described by the projection operator
\beq
P_s = \frac{1}{2} \left( 1 + s \hat Q \right),
\label{projs}
\eeq
where $s = \pm 1$. (Hats are used to denote operators only when necessary to distinguish from a classical counterpart).
For a system in initial state $\rho$ the probability for a single time measurement at time $t_1$  is
\beq
p_1(s) = {\rm Tr} \left( P_s (t_1) \rho \right),
\label{single}
\eeq
where $P_s (t)= e^{iHt} P_s e^{-iHt} $ is the projector in the Heisenberg picture and we use units in which $\hbar = 1$.
The probability for two sequential projective measurements at times $t_1, t_2$ is
\beq
p_{12} (s_1, s_2) = {\rm Tr} \left( P_{s_2} (t_2) P_{s_1} (t_1) \rho P_{s_1} (t_1) \right).
\label{p12}
\eeq
This matches $p_1 (s_1) $ when summed over $s_2$, but does not match the single time result
$ {\rm Tr} (P_{s_2} (t_2) \rho ) $ when summed over $s_1$.
Hence
the NSIT condition Eq.(\ref{NSIT}) is not in general satisfied in quantum mechanics, except for initial density operators diagonal in $\hat Q$ at time $t_1$.

It is also of interest to consider the two-time quasi-probability
\beq
q(s_1, s_2) = {\rm Re} {\rm Tr} \left( P_{s_2} (t_2) P_{s_1} (t_1) \rho \right),
\label{quasi}
\eeq
which is real and sums to $1$, but can be negative \cite{HalQ,GoPa}. It matches both single-time marginals $p_1 (s_1)$ and $p_2 (s_2)$ when summed over $s_2$ and $s_1$ respectively.
Following the useful general results of Refs.\cite{HaYe,Kly}, it has a very useful moment expansion,
\beq
q(s_1,s_2) = \frac {1}{4} \left(1 + \langle \hat Q_1 \rangle  s_1 +  \langle \hat Q_2 \rangle s_2  + C_{12} s_1 s_2 \right),
\label{mom}
\eeq
where the correlator is,
\beq
C_{12} = \half \langle \hat Q_1 \hat Q_2 + \hat Q_2 \hat Q_1 \rangle,
\label{corr2}
\eeq
and the averages are quantum-mechanical ones in these expressions.
Here, $\hat Q_i$ denotes the operator $\hat Q$ at time $t_i$, in parallel with the macrorealistic notation used in the Introduction.
(We will not use the notation $\hat Q(t)$ to denote time-dependence in order to avoid confusion with a notation used later in the many-valued case).
We see from Eq.(\ref{mom}) that, for a quantum-mechanical system, the two-time LG inequalities, Eq.(\ref{2time}), are equivalent to the conditions,
\beq
q(s_i, s_j) \ge 0. 
\eeq

The two-time measurement probability has a similar moment expansion
\beq
p_{12}(s_1,s_2) = \frac {1}{4} \left(1 + \langle \hat Q_1 \rangle  s_1 +   \langle \hat Q_2^{(1)}  \rangle s_2 
+ C_{12} s_1 s_2 \right).
\label{p12mom}
\eeq
Here,
\beq
\langle \hat Q_2^{(1)} \rangle = \langle \hat Q_2 \rangle + \half \langle [ \hat Q_1, \hat Q_2] \hat Q_1 \rangle
\label{Q2(1)}
\eeq
and has the interpretation as the average of $\hat Q_2$ in the presence of an earlier measurement at $t_1$ with the result summed out. The difference between Eq.(\ref{p12mom}) and the quasi-probability therefore vanishes when $ \langle \hat Q_2^{(1)} \rangle = \langle \hat Q_2 \rangle $
(for example, when $ \hat Q_1 $ and $ \hat Q_2 $ commute, but this is clearly not the only way). 
Note that the quasi-probability and the two-time measurement probability have the same correlation function \cite{Fri}.

We now introduce the description of interferences in relation to 
the quasi-probability Eq.(\ref{quasi}) and the two-time probability Eq.(\ref{2time}). This may be seen from the simple relationship (proved in greater generality in Appendix B), which is,
\beq
q(s_1, s_2) = p_{12} (s_1, s_2)  +\sum_{{s_1'} \atop {s_1' \ne s_1}}  \ {\rm  Re} D (s_1,s_2 |s_1',s_2),
\label{qp}
\eeq
where
\beq
D (s_1,s_2 |s_1',s_2) = {\rm Tr} \left(  P_{s_2} (t_2) P_{s_1} (t_1) \rho P_{s_1'} (t_1)  \right),
\label{DF1}
\eeq
is the so-called {\it decoherence functional} \cite{HalQ,GH2,Gri,Omn1,HalQIP}, and is a very useful quantity in what follows. The off-diagonal terms of the decoherence functional are measures of interference between the two different quantum histories represented by sequential pairs of projectors, $P_{s_2} (t_2) P_{s_1} (t_1)$ and $P_{s_2} (t_2) P_{s_1'} (t_1)$. For the dichotomic case considered here, there is in fact only one term in the sum over $s_1'$, namely $s_1' = - s_1$, so the interference terms, which for conciseness we denote $I (s_2)$, are simply
\beq
I(s_2) = {\rm Re} D(s_1,s_2|-s_1, s_2),
\eeq
and are independent of $s_1$.
Note that by inserting the moment expansions Eqs.(\ref{mom}), (\ref{p12mom}) in Eq.(\ref{qp}), we may make the identification $I(s_2) = \frac{1}{4}  ( \langle \hat Q_2 \rangle - \langle \hat Q_2^{(1)} ) s_2$. Hence the two-time interferences depend only on first moments, and not the two-time correlator.

When the off-diagonal terms are zero
there is no interference and we have $q(s_1,s_2) = p_{12} (s_1,s_2) $, and the NSIT condition Eq.(\ref{NSIT})
is satisfied exactly. However, noting that $p_{12}(s_1,s_2) $ is always non-negative, we see from Eq.(\ref{qp}) that 
the requirement that $q(s_1, s_2)$ is non-negative is equivalent to the following bounds on the off-diagonal terms of the decoherence functional:
\beq
- I(s_2)  \le p_{12} (s_1, s_2).
\label{ReDp}
\eeq
Hence, the two-time LG inequalities represent bounds on the degree of interference. 



From Eq.(\ref{DF1}), it is easy to see that the decoherence functional summed over $s_2$ gives zero for $s_1 \ne s_1'$, which means that $\sum_{s_2} I(s_2) = 0$. Hence 
the interference terms in Eq.(\ref{ReDp}) are fixed by just one independent quantity, which could be taken for example to be $I(+) = - I(-)$.
However, this single interference term has four upper bounds in Eq.(\ref{ReDp}), which is why there are, correspondingly, four two-time LG inequalities.
By contrast, the stronger NSIT condition Eq.(\ref{NSIT}) simply requires that all interference terms are zero. Since there is just one independent interference term the NSIT condition amounts to a single condition, $I(+) = 0$.

There is also a relationship between the interference terms $I(s_2)$ and the so-called ``coherence witness'' \cite{Wit}, which measures the degree to which the NSIT condition Eq.(\ref{NSIT}) is violated. We have
\bea
p_2 (s_2) - \sum_{s_1} p_{12} (s_1, s_2) &=& \sum_{{s_1, s_1'} \atop{s_1 \ne s_1'}}  \ {\rm  Re} D (s_1,s_2 |s_1',s_2),
\nonumber \\
&=& 2 I(s_2).
\label{Wit1}
\eea
Hence we identify the coherence witness with $2 I(s_2)$. (It is often defined with a modulus sign but the signed definition used here is convenient). 


In terms of measurements to check the two-time LG inequalities (or equivalently measure the quasi-probability Eq.(\ref{quasi})), the procedure is to carry out measurements of $ \langle \hat Q_1 \rangle$, $  \langle \hat Q_2 \rangle $ and $C_{12}$ in three 
separate experiments \cite{HalQ,HalLG4}.
These must be non-invasive (from a macrorealistic perspective) to meet the NIM requirement. This is trivial for the two averages (since only a single measurement is made in those experiments). For the case of $C_{12}$ this is accomplished using a pair of sequential measurements where the first one is an ideal negative measurement (in which the detector is coupled to, say, $Q_1=+1$ and a null result is taken to imply that $Q_1=-1$). This determines the sequential probability $p_{12} (s_1, s_2)$ from which the correlator is obtained (and as noted, coincides with the correlator in the quasi-probability $q(s_1,s_2))$.  
It should be added that the extent to which ideal negative measurements really meet the NIM requirement has been a matter of debate and alternative non-invasive measurement measurement protocols have been explored. (See for example Refs.\cite{HalNIM,HalCTVM} and references therein).

\subsection{LG Inequalities for Three Times}

A discussion of the relationship between the LG inequalities and bounds on the interferences following from Eq.(\ref{qp}) in the two-time case may also
be given for the three-time LG inequalities using the generalizations of the quasi-probability and two-time probability given in Appendix B. Following that discussion, we first consider the three-time quasi-probability
\beq
q(s_1, s_2, s_3) = {\rm Re} {\rm Tr} \left( C_{s_1 s_2 s_3} \rho \right)
\eeq
where 
\beq
C_{s_1 s_2 s_3} = P_{s_3} (t_3) P_{s_2} (t_2) P_{s_1} (t_1) .
\label{Csss}
\eeq
It has moment expansion,
\bea
q(s_1,s_2,s_3) = \frac{1}{8} ( 1 \! &+& s_1 \langle \hat Q_1 \rangle + s_2 \langle \hat Q_2 \rangle +s_3 \langle \hat Q_3 \rangle  \
\nonumber \\
&+&  s_1 s_2 C_{12} + s_2 s_3 C_{23} + s_1 s_3 C_{13} + s_1 s_2 s_3 {D}
),
\label{q123}
\eea
where $D$ is a triple correlator whose form is not needed here, and is not normally measured in LG experiments \cite{HalLG4,HaYe}.
Now we note that the quantum-mechanical expression for the correlators, Eq.(\ref{corr2}), are equivalently written,
\beq
C_{12} = \sum_{s_1 s_2 s_3} s_1 s_2 \ q (s_1, s_2, s_3),
\eeq
and likewise for $C_{23}$ and $C_{13}$.  We focus on the LG inequality Eq.(\ref{LG1}), which we write here as $L_1 \ge 0 $, where
\beq
L_1 = 1 + C_{12} + C_{13} + C_{23}.
\eeq
This can be written in the quantum case as
\beq
L_1 = \sum_{s_1 s_2 s_3} \left[  1 + s_1 s_2 + s_1 s_3 + s_2 s_3 \right] \ q( s_1, s_2, s_3).
\label{L1}
\eeq
Since the term in square brackets is always non-negative $L_1$ can only be negative if $q(s_1,s_2,s_3) < 0 $.

Eq.(\ref{pqd}) shows that
the quasi-probability $q(s_1,s_2,s_3)$ is related to the three-time sequential measurement probability $p_{123}(s_1,s_2,s_3)$ by
\beq
q(s_1,s_2,s_3) = p_{123} (s_1,s_2,s_3) + {\rm Re} {\rm Tr} \left(C_{s_1 s_2 s_3}\ \rho  \ \overline{C}_{s_1 s_2 s_3}^\dag \right),
\label{qpi}
\eeq
where $ \overline{C}_{s_1 s_2 s_3} = 1 - C_{s_1 s_2 s_3}$.
This means the interference terms need to be sufficiently large and negative in order for $q(s_1,s_2,s_3)$ to become negative, in parallel with the two-time case. However, unlike the two-time case, the LG inequality $L_1$ is a coarse-graining of the three-time quasi-probability and we need to be more specific in identifying the interference terms responsible for a LG violation.

A very convenient way of identifying the specific form of the interference terms in Eq.(\ref{qpi}) is to note that $p_{123} (s_1, s_2, s_3) $ also has a moment expansion, which is
\bea
p_{123} (s_1,s_2,s_3) = \frac{1}{8} ( 1 \! &+& s_1 \langle \hat Q_1 \rangle + s_2 \langle \hat Q_2^{(1)} \rangle +s_3 \langle \hat Q_3^{(12)} \rangle  \
\nonumber \\
&+&  s_1 s_2 C_{12} + s_2 s_3 C_{23}^{(1)} + s_1 s_3 C_{13}^{(2)} + s_1 s_2 s_3 {D}
).
\label{p123}
\eea
Here, as in Eq.(\ref{Q2(1)}), the superscripts on the averages and two-time correlators denote the presence of earlier or intermediate measurements whose results are summed over \cite{HalLG4}
(and the details of such averages and correlators is not relevant here -- we need only the general form of the interference terms). So for example, $C_{13}^{(2)} $ is the correlator for the time pair $t_1, t_3$ in the presence of an intermediate measurement at $t_2$ which has been summed out.
Taking the difference between Eq.(\ref{q123}) and Eq.(\ref{p123}), we find that the interference terms have the form
\bea
{\rm Re} {\rm Tr} \left(C_{s_1 s_2 s_3}\ \rho  \ \overline{C}_{s_1 s_2 s_3}^\dag \right)
= \frac{1}{8} \left(  s_2   (\langle \hat Q_2 \rangle - \langle \hat Q_2^{(1)} \rangle )  
+  s_3  (\langle \hat Q_3 \rangle - \langle \hat Q_3^{(12)} \rangle )  \right.
\nonumber \\
\left. + s_2 s_3 (C_{23} - C_{23}^{(1)}) + s_1 s_3 ( C_{13} - C_{13}^{(2)}) \right)
\eea
Inserting this in Eq.(\ref{qpi}) and Eq.(\ref{L1}), we find
\beq
L_1 =  \sum_{s_1 s_2 s_3} \left[  1 + s_1 s_2 + s_1 s_3 + s_2 s_3 \right] \ p_{123}( s_1, s_2, s_3)
+  (C_{23} - C_{23}^{(1)}) +  ( C_{13} - C_{13}^{(2)}),
\eeq
and note that the first order moment terms have dropped out entirely. The first term on the right-hand side is clearly non-negative, and hence the interference terms responsible for a three-time LG violation are the remaining terms on the right, involving the difference between two-time correlators.  These terms are completely independent of the quantities controlling the two-time LG violations, which as we saw above, are the quantities  $(\langle \hat Q_2 \rangle - \langle \hat Q_2^{(1)} \rangle ) s_2  $ (and similarly for the other two-time pairs). Hence the two-time and three-time LG inequalities test for the presence of completely independent types of interference terms.
This observation is relevant to the study of L\"uders bound violations in Section 6.

We briefly note that the three-time LG inequalities Eqs.(\ref{LG1})--(\ref{LG4}) may be written in terms of the quasi-probability Eq.(\ref{quasi}), as
\beq
q(s_1,-s_2) + q(s_2,-s_3) + q(-s_1,s_3) \le 1.
\label{LG3q}
\eeq
We will see that this is a convenient starting point for generalizations. 





\section{Conditions for Macrorealism using Leggett-Garg Inequalities at Two Times}

We now come to the main work of this paper which is to establish MR conditions for many-valued variables. In this Section and the next we do so using LG inequalities, which, in the language of Appendix A, characterise weak MR. In this Section we focus on two times and consider three and more times in the next Section.

\subsection{Projectors for Many-Valued Variables}

We suppose that measurements on our system may be described by a set of projection operators $E_n$, where $n=1, 2, \cdots N$ and $\sum_n E_n = \id$, where $\id$ denotes the identity operator.
(To avoid confusion, we use the notation $P_s$ with $s= \pm 1$, for projectors only in the dichotomic case).
These could be fine-grained measurements $E_n = |n \rangle \langle n |$ on an $N$-level system, or coarse-grained measurements on a system of dimension greater than $N$. These projections may be regarded as measurements of any hermitian operator which has a spectral expansion in terms of the $E_n$,
and this is the sort  of ``many-valued variable" we have in mind, but we will not make explicit use of such an operator in what follows.
Some of the MR conditions we develop in what follows may be extended to the case in which the $E_n$ describe ambiguous measurements (see Ref.\cite{Ema} for example) but we will not spell this out explicitly here.

As stated in the Introduction, our strategy for deriving MR conditions is to link the $E_n$ with a set of dichotomic variables.
The widest class of such variables have the form,
\beq
\hat Q = \sum_{n=1}^N \epsilon (n) E_n,
\label{Qdef}
\eeq
where $\epsilon (n) $ takes values $\pm 1$ with at least one $+1$ and one $-1$. However,
for our purposes we have found that it is sufficient in almost all cases to focus on a more restricted class, consisting of the $N$ dichotomic variables in which $\epsilon (n)$ has a single $+1$ and the rest of the values are $-1$. (The only exception we have encountered concerns NSIT conditions for $N > 3$ briefly discussed later on).
Each $Q$ therefore has the form, for fixed $n$,
\beq
\hat Q(n) = E_n - \bar E_n.
\eeq
where $\bar E_n = \id - E_n$, the negation of $E_n$. Each projector $E_n$ may therefore be written,
\beq
E_n = \frac{1}{2} \left( \id + \hat Q(n) \right)
\label{PQn}
\eeq
Since the $E_n$ sum to the identity, the $N$ dichotomic variables $\hat Q(n)$ satisfy
\beq
\sum_{n=1}^N \hat Q(n) = (2 - N) \id
\label{QN}
\eeq
This means that they are still more than the minimum needed to uniquely fix the state of the system. This non-minimal set of variables is convenient to use but it will sometimes be convenient to revert to a set of $N-1$ independent variables.

So far the description is quantum-mechanical but since LG conditions are best formulated in a macrorealistic setting, we introduce the classical analogues of the projectors $E_n$, denoted $\P (n)$, which take values $0$ or $1$, and a corresponding classical object $Q(n)$, with the two related by
\beq 
\P (n) = \frac{1}{2} \left( 1 + Q(n) \right).
\eeq

In what follows we are interested in the values of $Q(n)$ and $\hat Q(n)$ at times $t_i$ and, following the notation introduced earlier, we denote those values by $Q_i (n)$ and $\hat Q_i (n)$.
The label $n$ always denotes one of the $N$ dichotomic variables and times are always denoted using subscripts. (This is why we avoid the commonly-used notation $Q(t)$ to denote time-dependence, as noted earlier).

\subsection{Two-Time LG Inequalities}

We now consider conditions for weak MR at two-times for many-valued variables, using generalizations of the two-time LG inequalities for the dichotomic case Eq.(\ref{2time}). We suppose that non-invasive measurements are made, as described in Section 2, on the $N$ dichotomic variables $Q(n)$ which determine the averages $\langle Q_1 (n_1) \rangle $, $\langle Q_2 (n_2) \rangle$ and the correlator $\langle Q_1 (n_1) Q_2 (n_2) \rangle $. From these quantities we seek to construct a candidate probability $p(n_1,n_2)$. It is reasonably clear that the desired probability is
\beq
p(n_1, n_2) =\langle \P_2 (n_2) \P_1 (n_1) \rangle,
\label{LGN2}
\eeq
where again the subscripts are time labels.
This is clearly the classical analogue of  the quantum-mechanical quasi-probability,
\beq
q(n_1, n_2) = {\rm Re} {\rm Tr} \left( E_{n_2} (t_2) E_{n_1} (t_1) \rho \right),
\label{quasin}
\eeq
which is the natural generalization of Eq.(\ref{quasi}). The desired two-time LG inequalities are then simply the requirement that the two-time probability is non-negative, which, written out in terms of the $Q(n)$ variables, read:
\beq
1 + \langle Q_1 (n_1) \rangle + \langle Q_2 (n_2) \rangle + \langle Q_1 (n_1) Q_2 (n_2) \rangle \ge 0.
\label{LGN2Q}
\eeq
These $N^2$ LG inequalities are necessary and sufficient conditions for MR at two times. Necessity is trivially established and sufficiency follows since the probabilities themselves, $p(n_1,n_2)$, are proportional to the inequalities.
The LG inequalities involve a set of averages and correlators of the $N$ variables $Q(n)$. 
However, as indicated, this is a non-minimal set since they satisfy Eq.(\ref{QN}). This means that in practice it is only necessary to make measurements on $N-1$ of the $Q(n)$, and the averages and correlators involving the unmeasured variable readily obtained using Eq.(\ref{QN}). We will see this explicitly below for the $N=3$ case.


It may seem unusual that Eq.(\ref{LGN2Q}) contains only plus signs, in contrast to the usual two-time LG inequalities Eq.(\ref{2time}), which contains plus and minus signs. However, this relates to the non-minimal nature of the set of variables $Q(n)$ and different forms of the LG inequalities may be obtained by taking linear combinations. In particular, suppose Eq.(\ref{LGN2}) is summed over all $n_1 \ne n_1'$ for some $n_1'$, and we use the fact that
\beq
\sum_{n_1 \ne n_1'} \P_1 (n_1) = 1 - \P_1 (n_1') \equiv \overline{ \P}_1 (n_1').
\label{PPbar}
\eeq
This is easily seen to have the effect of replacing $Q_1(n_1)$ with $ -Q_1(n_1')$, so we get,
\beq
1 - \langle Q_1 (n_1') \rangle + \langle Q_2 (n_2) \rangle - \langle Q_1 (n_1') Q_2 (n_2) \rangle \ge 0.
\label{minus}
\eeq
This means that the set of inequalities Eq.(\ref{LGN2Q}) is equivalent to any other set in which any of the $Q(n)$'s have reversed sign. It is therefore sufficient to work with any one set with a fixed set of signs. 
Note also that for the dichotomic case, $N=2$, Eq.(\ref{QN}) implies that the two dichotomic variables $Q(1)$, $Q(2)$ satisfy $Q(1) = - Q(2)$,  and the four LG inequalities Eq.(\ref{LGN2Q}) coincide with the two-time LG inequalities Eq.(\ref{2time}), as required.


\subsection{Two-Time LG Inequalities for the $N=3$ Case}

The LG inequalities Eq.(\ref{LGN2Q}) are simplest in form when written in terms of the non-minimal set of $N$ variables $Q(n)$, but in terms of measurements to check them, it is actually only necessary to measure $N-1$ variables at each time and use Eq.(\ref{QN}) to determine the correlators and averages for the remaining unmeasured variable. We write this out explicitly for the case $N=3$. It is notationally more convenient in the $N=3$ case to take $n=A,B,C$ (following common usage \cite{Ema}).  Instead of the three variables  $Q(n)$ for $n=A,B,C$, for notational convenience we use the three dichotomic variables $Q$, $R$, $S$ and use the classical projectors 
\bea
\P (A) & =& \frac{1}{2}(1+Q), \\ 
\P(B) &=& \frac{1}{2}(1+R), \\
\P(C) &=& \frac{1}{2} (1+S). 
\eea
The requirement that these sum to $1$ is equivalent to the statement $ Q+R+S = -1$.
This relation between $Q$, $R$ and $S$ may seem to unduly constrain the system (because for example we cannot have $Q=R=+1$) but this does not matter since at each time only one or the other is measured.

In terms of these variables the nine two-time LG inequalities have the form Eq.(\ref{LGN2Q}) and  four of those relations are:
\bea
1 +   \langle Q_1 \rangle +  \langle Q_2 \rangle +  \langle Q_1 Q_2 \rangle  & \ge & 0,
\label{LGNa}
\\
1 +   \langle R_1 \rangle +  \langle Q_2 \rangle +  \langle R_1 Q_2 \rangle  & \ge & 0, 
\\
1 +    \langle Q_1 \rangle +  \langle R_2 \rangle +  \langle Q_1 R_2 \rangle  & \ge & 0, 
\\
1 +    \langle R_1 \rangle +  \langle R_2 \rangle +  \langle R_1 R_2 \rangle  & \ge & 0.
\label{LGNd}
\eea
The other five are similar in form and all involve the variable $S$. However, the relation $Q+R+S=-1$ means that all averages and correlators involving $S$ may be expressed in terms of averages and correlators involving $Q$ and $R$. Carrying this out explicitly yields:
\bea
\langle Q_1 Q_2 \rangle + \langle Q_1 R_2 \rangle + \langle R_1 Q_2 \rangle + \langle R_1 R_2 \rangle & \ge & 0,
\\
\langle Q_1 \rangle + \langle R_1 \rangle + \langle Q_1 Q_2 \rangle + \langle R_1 Q_2 \rangle & \le & 0,
\\
\langle Q_1 \rangle + \langle R_1 \rangle + \langle Q_1 R_2 \rangle + \langle R_1 R_2 \rangle & \le & 0,
\\
\langle Q_2 \rangle + \langle R_2 \rangle + \langle Q_1 Q_2 \rangle + \langle Q_1 R_2 \rangle & \le & 0,
\\
\langle Q_2 \rangle + \langle R_2 \rangle + \langle R_1 Q_2 \rangle + \langle R_1 R_2 \rangle & \le & 0.
\label{LGNi}
\eea
Note that the last four inequalities are upper, not lower bounds. This form is the most useful form to use for experimental tests since no measurements on $S$ are required, only on $Q$ and $R$. This will clearly be true more generally -- measurements on only $N-1$ of the $Q(n)$ variables are required.

An experiment to test inequalities of this type on a three-level system was in fact carried out recently \cite{Ema,EmINRM}. The experiment used ambiguous measurements to determine all nine components of the quasi-probability Eq.(\ref{quasin}) and observed violations. It therefore represents a test of a complete set of necessary and sufficient conditions for MR at two times for a three-level system. (This experiment also involved a NSIT condition, discussed further below).





\section{Conditions for Macrorealism using Leggett-Garg Inequalities at Three and More Times }

We turn now to the three-time case and look for conditions for weak MR involving set of three-time LG inequalities for many-valued variables in conjunction with the two-time LG inequalities already derived. We thus generalize the results of Refs.\cite{JHFine,HalLG4,HaMa}.

\subsection{Three-Time LG Inequalities and Fine's Theorem}

We seek necessary and sufficient conditions for the existence of a joint probability $p(n_1,n_2,n_3)$ matching the three pairwise probabilities  $p(n_1,n_2)$, $p(n_2,n_3)$ and $p(n_1,n_3)$ which take the form $ \langle \P_i(n_i) \P_j(n_j) \rangle$. 
The dichotomic case expressed in the form Eq.(\ref{LG3q}) suggests
the following proposition: the necessary and sufficient conditions are a set of two-time LG inequalities of the form $ \langle \P_i(n_i) \P_j(n_j) \rangle \ge 0 $ ensuring the non-negativity of the pairwise probabilities, together with 
the $N^3$ inequalities,
\beq
\langle \P_1 (n_1) \overline{\P}_2 (n_2) \rangle
+ \langle \P_2 (n_2) \overline{\P}_3 (n_3) \rangle
+ \langle \overline{\P}_1 (n_1) \P_3 (n_3) \rangle \le 1.
\eeq
Writing $ \P (n) = \frac{1}{2} (1 + Q(n) ) $, these inequalities read
\beq
1+  \langle Q_1(n_1) Q_2(n_2)  \rangle + \langle Q_2(n_2) Q_3(n_3)  \rangle + \langle Q_1(n_1) Q_3(n_3) \rangle \ge 0.
\label{LGN3Q}
\eeq
We will prove the proposition below.
Although this is a complete set of LG inequalities as it stands, like the two-time case it appears to involve only one set of signs in front of the correlators. However, again it is possible to demonstrate equivalence with LG inequalities with another set of signs. This can be accomplished for example, by summing Eq.(\ref{LGN3Q}) over all $n_1 \ne n_1'$ (as in Eqs.(\ref{PPbar}), (\ref{minus})), and using Eq.(\ref{QN}). This yields a sum of a three-time LG inequalities of the form Eq.(\ref{LGN3Q}) with $Q_1 (n_1) $ replaced with $ - Q_1 (n_1) $ and a two-time LG inequality. Hence taken together with the two-time LG inequalities (which we assume hold), the three-time LG inequalities may be written in a number of different forms.

For $N=2$, we may write $Q(1) = Q$ and $Q(2) = - Q$, and we see that the eight inequalities Eq.(\ref{LGN3Q}) boil down to the four familiar LG inequalities Eqs.(\ref{LG1})-(\ref{LG4}). This simplification does not seem to happen for $N \ge 3$. For $N=3$, for example, we have twenty-seven three-time LG inequalities in terms of the $Q(n)$. However, the fact that the $Q(n)$ are a non-minimal set of dichotomic variables, obeying Eq.(\ref{QN}), reduces the number of correlators that have to be measured, as we saw in the two-time case already. In particular, we may write out the twenty-seven LG inequalities in the $N=3$ case in terms of the variables $Q$, $R$, $S$ introduced in Section 3(C), satisfying $Q+R+S=-1$. Any correlators or averages involving $S$ may then be expressed in terms of those involving only $Q$ and $R$. We thus obtain a set of three-time LG inequalities analogous to the two-time set, Eqs.(\ref{LGNa})-(\ref{LGNi}). These are written out in full in Appendix C.


\subsection{Proof of the Generalized Fine's Theorem}

We now prove the above proposition. Necessity trivially follows from the simple identity,
\beq
\langle ( Q(n_1) + Q(n_2) + Q(n_3) )^2  \rangle \ge 1.
\eeq
To prove sufficiency,
we use the moment expansion of $p(n_1,n_2,n_3)$, making use of Eq.(\ref{PQn}),
defined by,
\beq
\begin{split}
p(n_1,n_2,n_3) = \langle  \P_1 ( & n_1)   \P_2 (n_2) \P_3 (n_3) \rangle
\\
= \frac{1}{8} \bigg( 1 &+ \langle Q_1(n_1) \rangle + \langle Q_2(n_2) \rangle + \langle Q_3(n_3) \rangle
\\
&+  \langle Q_1(n_1) Q_2(n_2)  \rangle + \langle Q_2(n_2) Q_3(n_3)  \rangle + \langle Q_1(n_1) Q_3(n_3) \rangle
\\
&+ \langle Q_1(n_1) Q_2(n_2) Q_3(n_3) \rangle
\bigg).
\end{split}
\label{pn3}
\eeq
This clearly matches the pairwise marginals as required. It involves a set of unfixed triple correlators $ \langle Q_1(n_1) Q_2(n_2) Q_3(n_3) \rangle$ so the aim is to determine the conditions under which these may be chosen to ensure that 
\beq
0 \le p(n_1, n_2, n_3) \le 1.
\label{UL}
\eeq
Since the probability $p(n_1,n_2,n_3)$ sums to $1$ by construction, the upper bound of $1$ is guaranteed as long as we can show that all $N^3$ terms are non-negative. 
However, it turns out to be more convenient to focus on fixed values of $n_1, n_2, n_3$ and prove that $p(n_1,n_2,n_3)$ satisfies Eq.(\ref{UL}). The lower bound is clearly satisfied by suitable choice of the triple correlator.
It turns out that the upper bound is most easily handled by re-expressing it as a set of lower bounds on related probabilities, using the identity,
\beq
\langle \left( \P_1 (n_1) + \overline{\P}_1 (n_1) \right)  \left( \P_2 (n_2) + \overline{\P}_2 (n_2) \right) 
 \left( \P_3 (n_3) + \overline{\P}_3 (n_3) \right) 
\rangle = 1.
\eeq
When expanded out, this is readily seen to imply that,
\bea
1 - p(n_1, n_2, n_3) &=&
\langle   \P_1 ( n_1)   \P_2 (n_2) \overline{\P}_3 (n_3) \rangle
+\langle  \P_1 ( n_1)  \overline{ \P}_2 (n_2) \P_3 (n_3) \rangle
\nonumber \\
&+&\langle  \overline{\P}_1 (n _1)   \P_2 (n_2) \P_3 (n_3) \rangle
+\langle  \overline{\P}_1 ( n_1)   \overline{\P}_2 (n_2) \overline{\P}_3 (n_3) \rangle
\nonumber \\
&+&\langle  \overline{\P}_1 ( n_1)   \overline{\P}_2 (n_2) \P_3 (n_3) \rangle
+\langle  \overline{\P}_1 ( n_1)   \P_2 (n_2)\overline{\P}_3 (n_3) \rangle
\nonumber \\
&+&\langle  \P_1 ( n_1)   \overline{\P}_2 (n_2) \overline{\P}_3 (n_3) \rangle.
\label{negp}
\eea
It follows that $p(n_1,n_2,n_3) \le 1$ as long as the seven probabilities on the right-hand side are non-negative.
Each of these probabilities has a moment expansion of the form Eq.(\ref{pn3}), in which one, two or three of the $Q(n)$'s have their sign flipped, and which are readily seen to yield more upper and lower bounds on the triple correlator.

The remaining steps in the proof are very similar to the dichotomic case covered in Refs.\cite{JHFine,HaMa}.
For simplicity we write $p(n_1,n_2,n_3) \ge 0 $ as,
\beq
F (Q_1 (n_1), Q_2 (n_2), Q_3 (n_3) )  + \langle Q_1(n_1) Q_2(n_2) Q_3(n_3) \rangle  \ge 0,
\eeq
where $ F$ is read off from Eq.(\ref{pn3}). This clearly gives a lower bound on the triple correlator. 
The triple correlator has the same sign for the moment expansion of the last three probabilities on the right-hand side of Eq.(\ref{negp}). We thus find that non-negativity of four of the probabilities is ensured if
the triple correlator has a total of four lower bounds defined  by
\beq
 \langle Q_1(n_1) Q_2(n_2) Q_3(n_3) \rangle \ \ge \ -F ( s_1 Q_1(n_1), s_2 Q_2(n_2), s_3 Q_3(n_3))  \bigg|_{s_1 s_2 s_3 = +1},
\eeq
where the $s_i$ take values $\pm 1$.
Similarly, the first four probabilities on the right-hand side of Eq.(\ref{negp}), have a minus sign in front of the triple correlator, and we thus obtain the four upper bounds defined by,
\beq
F ( s_1 Q_1(n_1), s_2  Q_2(n_2), s_3  Q_3(n_3))  \bigg|_{s_1 s_2 s_3  = -1}\  \ge \ \langle Q_1(n_1) Q_2(n_2) Q_3(n_3) \rangle.
\eeq
A triple correlator ensuring that Eq.(\ref{UL}) holds
may therefore be found as long as the four lower bounds are less than the four upper bounds, i.e.
\beq
F ( s_1' Q_1(n_1), s_2'  Q_2(n_2), s_3'  Q_3(n_3))  \bigg|_{s_1' s_2' s_3'  = -1}
+ F ( s_1 Q_1(n_1), s_2 Q_2(n_2), s_3 Q_3(n_3))  \bigg|_{s_1 s_2 s_3 = +1} \ge 0,
\eeq
for all possible choices of $s_i, s_i'$ satisfying the stated restrictions.
Written out in full, these inequalities are readily seen to coincide with  the two-time LG inequalities Eq.(\ref{LGN2Q}) and the three-time LG inequalities Eq.(\ref{LGN3Q}) (and their variants obtained under sign changes). 
One also needs to check that the choice of triple correlator satisfying these upper and lower bounds lies in the correct range, $[-1,1]$, however this is also ensured by the two and three-time LG inequalities (and this is readily seen from the proof in Ref.\cite{HaMa}). This proves sufficiency.





\subsection{Four and More Times}

The dichotomic case is usually formulated in terms of measurements at both three and four times. It was recently generalized to measurements made at an arbitrary number of times and the corresponding Fine's theorem derived \cite{HaMa}. This extension from three to many times made use of a generalization of  a famous ansatz of Fine, which expresses the four-time problem entirely in terms of three-time probabilities. This ansatz was originally given in terms of dichotomic variables, but we make the simple observation that the ansatz still works for many-valued variables and, for the four-time case is:
    \beq
    p(n_1, n_2, n_3,n_4)=\frac{p(n_1, n_2, n_3)\ p(n_1, n_3, n_4)}{p(n_1, n_3)}.
    \label{fineansatz}
    \eeq
This is the solution to the matching problem in which we seek a probability $   p(n_1, n_2, n_3,n_4)$ matching the four pairwise marginals, $p(n_1,n_2)$, $p(n_2,n_3)$, $p(n_3,n_4)$ and $p(n_1,n_4)$. It reduces the problem of showing that the four-time probability is non-negative to that of showing that two three-time probabilities are non-negative, which is guaranteed if two appropriate sets of three-time LG inequalities are satisfied, which we may choose to be,
\bea
1 + \langle Q(n_1) Q(n_2) \rangle + \langle Q(n_2) Q(n_3) \rangle + \langle Q(n_1) Q(n_3) \rangle  & \ge & 0, 
\label{QQ123}
\\
1 - \langle Q(n_1) Q(n_3) \rangle + \langle Q(n_3) Q(n_4) \rangle - \langle Q(n_1) Q(n_4) \rangle  & \ge & 0.
\label{QQ134}
\eea
The choice of signs here is purely for convenience and exploits the fact equivalence between different sets of three-time LG inequalities under sign flips of the $Q(n)$. Eliminating the unfixed correlator $ \langle Q(n_1) Q(n_3) \rangle $ between these two inequalities we obtain the $N^4$ inequalities:
\beq
\langle Q(n_1) Q(n_2) \rangle +  \langle Q(n_2) Q(n_3) \rangle + \langle Q(n_3) Q(n_4) \rangle -\langle Q(n_1) Q(n_4) \rangle \ge -2
\label{LG41}
\eeq
We compare this with the set of CHSH-type inequalities, which for fixed $n_i$ consists of eight inequalities, the first two of which are,
\beq
-2 \le   \langle Q(n_1) Q(n_2) \rangle 
 + \langle Q(n_2) Q(n_3) \rangle 
 + \langle Q(n_3) Q(n_4) \rangle 
  -\langle Q(n_1) Q(n_4) \rangle \le 2,
\label{CHSH}
\eeq
and the remaining three pairs are obtained by moving the minus sign to the other three possible locations. The condition Eq.(\ref{LG41}) just derived is clearly just one of these. However, again using the non-minimal property of the set of $Q(n)$ and the consequent possibility of 
sign flips described in Eqs.(\ref{PPbar}), (\ref{minus}),
any one of these eight inequalities may be transformed into any other. Hence the necessary and sufficient conditions for (weak) MR at four times, consist of the non-negativity of the pairwise marginals together with any one of the CHSH-type inequalities 
Eq.(\ref{CHSH}). As expected these conditions reduce to the standard set of eight four-time LG inequalities in the dichotomic case with $Q(1) = - Q(2) =  Q$.

The complete set of LG inequalities for arbitrarily many times in the dichotomic case was given in Ref.\cite{HaMa}. This may be generalized to the case of many-valued variables at many times by proceeding along the same lines as the four-time case above.

\section{No-Signaling in Time Conditions}

We now consider stronger conditions for MR characterized by NSIT conditions.
We focus primarily on the two-time case, with brief mention of three-time MR conditions.

\subsection{Two-Time NSIT Conditions}

The NSIT condition Eq.(\ref{NSIT}) natural generalizes to the $N$ conditions,
\beq
\sum_{n_1 = 1}^N p_{12} (n_1, n_2) = p_2 (n_2).
\label{NSITn}
\eeq
Since both sides sum to $1$, the number of independent conditions is $N-1$.
However, for $N \ge 3$, this is not the only type of NSIT condition. 
One can instead measure any one of a number of dichotomic variables $Q$ at the first time to determine a probability $p^Q_{12} (s_1, n_2) $ for $s_1 = \pm 1 $, to which there corresponds a NSIT condition,
\beq
\sum_{s_1 } p^Q_{12} (s_1, n_2) = p_2 (n_2).
\label{NSITs}
\eeq
There will in general be a number of conditions of this type, depending on how $Q$ is defined. Recalling that the NSIT condition Eq.(\ref{NSIT}) is only satisfied for zero interference, the natural way to determine the most complete set of NSIT conditions in the $N \ge 3$ case is to look at the interferences, generalizing the analysis of Section 2, and derive a set of NSIT conditions which ensure that they are all zero. (Unlike the analysis of LG inequalities presented above, the analysis of NSIT conditions is quantum-mechanical in nature. A macrorealistic presentation is still possible here but we find the quantum-mechanical ones most convenient since it can take advantage
of the machinery of the decoherent histories approach briefly outlined in Appendix B).

The probability $p^Q_{12} (s_1, n_2)$ cannot simply be obtained from $p_{12} (n_1, n_2)$ by coarse graining over $n_1$ because of interferences. As described in Appendix B, both probabilities and interference terms are described by the decoherence functional,
\beq
D (n_1,n_2 |n_1',n_2) = {\rm Tr} \left(  E_{n_2} (t_2) E_{n_1} (t_1) \rho E_{n_1'} (t_1)  \right),
\label{DFN}
\eeq
and both of the probabilities $p_{12}(n_1,n_2)$ and $p_{12}^Q (s_1, n_2)$ may be obtained from it, as we now show.
The interferences themselves are represented by the off-diagonal terms of the decoherence functional.
The probability $p_{12} (n_1, n_2)$ is simply $D(n_1,n_2|n_1,n_2)$. The probability $p^Q_{12} (s_1, n_2)$  is the diagonal part of the decoherence functional,
\beq
D (s_1,n_2 |s_1',n_2) = {\rm Tr} \left(  E_{n_2} (t_2) P_{s_1} (t_1) \rho P_{s_1'} (t_1)  \right),
\label{DFS}
\eeq
where $P_s$ is the projector onto the values of $Q$, Eq.(\ref{projs}), and may be written in terms of the projectors $E_n$ as,
\beq
P_s = \sum_{n} c_{sn} E_n.
\label{csn}
\eeq
Here, the coefficients $c_{sn}$ are $0$ or $1$ and depend in a simple way on how $Q$ is defined. We thus see that these two decoherence functionals are related by
\beq
D (s_1,n_2 |s_1',n_2) = \sum_{n_1, n_1'} c_{s_1 n_1} c_{s_1' n_1'} D (n_1,n_2 |n_1',n_2).
\eeq
This, via Eq.(\ref{DFS}), gives an expression for  $p^Q_{12} (s_1, n_2)$ in terms of Eq.(\ref{DFN}).

The NSIT conditions in a quantum-mechanical setting may be checked by writing each side in terms of the decoherence functional.
Consider first Eq.(\ref{NSITn}). We have
\bea
p_2 (n_2) &=&  \sum_{n_1, n_1'} D (n_1,n_2 |n_1',n_2),
\nonumber \\
&=& \sum_{n_1} p_{12} (n_1, n_2) + \sum_{{n_1, n_1'} \atop {n_1 \ne n_1'}}  {\rm Re} D (n_1,n_2 |n_1',n_2).
\nonumber \\
\label{inter}
\eea
We again introduce a slightly more condensed notation for the interference terms,
\beq
I_{n_1 n_1'} (n_2)  = {\rm Re} D (n_1,n_2 |n_1',n_2),
\label{Idef}
\eeq
for $n_1 \ne n_1'$.
The coherence witness for the NSIT condition Eq.(\ref{NSITn}) is a sum terms of this form, so
the NSIT condition is violated unless the interference terms are zero.

Similarly, the violation of Eq.(\ref{NSITs}) is seen to be,
\bea
p_2 (n_2) &=&  \sum_{s_1 s_1'} D (s_1,n_2 |s_1',n_2),
\nonumber \\
&=& \sum_{s_1} p^Q_{12} (s_1, n_2) + \sum_{ {s_1,s_1'} \atop {s_1 \ne s_1'}} {\rm Re} D (s_1,n_2 |s_1',n_2),
\nonumber \\
&=&  \sum_{s_1} p^Q_{12} (s_1, n_2) + \sum_{{s_1,s_1'} \atop {s_1 \ne s_1'}}  \sum_{n_1, n_1'} c_{s_1 n_1} c_{s_1' n_1'} I_{n_1 n_1'} (n_2).
\label{NSITsV}
\eea
This shows that the two different types of NSIT conditions, Eq.(\ref{NSITn}) and Eq.(\ref{NSITs}), imply that different combinations of interference terms are zero. It then seems reasonably clear that we can make all the interference terms zero, as required by choosing suitable combinations of NSIT conditions. This is most easily seen in specific examples.


For general $N$ there are $ N(N-1)/2$ ways in which $n_1 \ne n_1'$. The interferences terms sum to zero when summed over $n_2$ which means that are $N-1$ interferences terms for fixed $n_1 \ne n_1'$, and therefore a total of $ N (N-1)^2/2$ independent interference terms. This means there is just one for the $N=2$ case, as we know already, but this jumps up to six for the $N=3$ case, which we now study in detail.

\subsection{Two-Time NSIT Conditions for the $N=3$ Case}

For the $N=3$ case the violation of the NSIT condition Eq.(\ref{NSITn}) reads,
\beq
p_2 (n_2) = \sum_{n_1} p_{12} (n_1, n_2)  + 2 I_{AB} (n_2) + 2  I_{AC} (n_2) + 2 I_{BC} (n_2),
\label{conA}
\eeq
where we use $I_{n_1 n_1'} (n_2)$ defined in Eq.(\ref{Idef}) and again use the labelling $n=A,B,C$ for the $N=3$ case.
Each interference term sums to zero when summed over $n_2$, so each term $I_{n_1 n_1'} (n_2)$ has two independent values for fixed $n_1, n_1'$. Hence there are a total of six interference terms as expected. There are three different choices for the NSIT condition Eq.(\ref{NSITs}) depending on how the dichotomic variables $Q(n)$ are defined.
We denote the three choices $Q,R,S$, where $ Q = E_A - \bar E_A$, $ R = E_B - \bar E_B$ and $S = E_C - \bar E_C $,
So for example, for $Q$, the values of $c_{sn}$ defined in Eq.(\ref{csn}) are $c_{+,A} = c_{-,B} = c_{-,C} = 1$ and the rest zero.
From Eq.(\ref{NSITsV}), this yields the NSIT violations,
\bea
p_2 (n_2) &=& \sum_{s_1} p^Q_{12} (s_1, n_2)  + 2  I_{AB} (n_2) + 2 I_{AC} (n_2),
\label{conB}
\\
p_2 (n_2) &=& \sum_{s_1} p^R_{12} (s_1, n_2)  + 2 I_{AB} (n_2) + 2 I_{BC} (n_2),
\label{conC}
\\
p_2 (n_2) &=& \sum_{s_1} p^S_{12} (s_1, n_2)  + 2 I_{AC} (n_2) + 2 I_{BC} (n_2).
\label{conD}
\eea
Eqs.(\ref{conA})-(\ref{conD}) is a set of eight conditions (since each one involves two independent conditions for $N=3$), but since there are only six interference terms to be constrained to zero, clearly only six of the eight NSIT conditions are required. For example, we could require that the interference terms in Eqs.(\ref{conB})-(\ref{conD}) vanish, which implies that all of the $I_{n_1 n_1'} (n_2)$ are zero. Or, we could require that the interference terms in Eq.(\ref{conA}) vanish along with the interferences terms in any pair of Eqs.(\ref{conB})-(\ref{conD}), with the same consequence.


We now note a significant new  and general feature that arises for $N \ge 3$, which is that the hierarchical relationship between two-time LG inequalities and NSIT conditions outlined in Appendix A for the dichotomic case becomes more complicated.
In the dichotomic case, the relationship between these two different types of conditions is simple. As we can see from Eq.(\ref{qp}), if $p_{12} (s_1, s_2)$ satisfies NSIT, then the interference term vanishes and $q(s_1, s_2) = p_{12} (s_1, s_2)$ and is therefore non-negative, i.e. the two-time LG inequalities hold. Equivalently, if the LG inequalities fail, NSIT must be violated. (However, the LG inequalities may still hold even if NSIT is violated). 

In the many-valued case, if a suitably large set of independent NSIT conditions are satisfied then all possible interference terms vanish. And since
Eq.(\ref{pqd}) for the many-valued case reads,
\beq
q(n_1, n_2) = p_{12} (n_1, n_2)  + \sum_{{n_1'} \atop {n_1' \ne n_1}} \ I_{n_1 n_1'} (n_2),
\label{qpN}
\eeq
this means that $q(n_1,n_2) \ge 0 $, i.e. the two-time LG inequalities hold. So far this hierarchical relationship 
is the same as the dichotomic case, as outlined in Appendix A.
However, we can have a situation in which some, but not all, combinations of the interference terms vanish. This would mean that the LG inequalities could be violated but some of the NSIT conditions are still satisfied.

For example in the $N=3$ case, consider the NSIT conditions Eq.(\ref{NSITn}) along with the LG inequalities for the dichotomic variable $Q = E_A - \bar E_A $. It is then readily shown that the quasi-probability (i.e. the set of two-time LG inequalities) is,
\bea
q(s_1, s_2) = p_{12}^Q (s_1, s_2) +  I_{AB} (s_2) +  I_{AC} (s_2),
\label{LGqpI}
\eea
where $I_{AB} (+) = I_{AB} (A)$ and $I_{AB}(-) = I_{AB}(B) + I_{AB} (C) $ and similarly for $I_{AC} (s_2)$. It is then clearly  possible for these interference terms to be non-zero and sufficiently negative that $q(s_1, s_2) \le 0 $ (i.e. the two-time LG inequalities fail), but with the sum over interference terms in Eq.(\ref{conA}) equal to zero, so that the two NSIT conditions Eq.(\ref{NSITn}) are satisfied.

Precisely such a situation was observed in two recent experiments which both note two-time LG violations when NSIT conditions are satisfied in a three-level system \cite{Ema,EmINRM,3box}. 
This seems surprising on the face of it, but these papers consider only NSIT conditions of the form Eq.(\ref{NSITn}). If a complete set of all six NSIT conditions is imposed then all interferences are zero and all two-time LG inequalities must be satisfied.
A similar feature was observed in a recent LG analysis of the triple slit experiment \cite{HalDS}.

Note that there are still {\it some} hierarchical relationships between NSIT conditions and LG inequalities. For example, if the NSIT time condition involving $Q$ is satisfied, i.e. the interference terms in Eq.(\ref{conB}) vanish, then the quasi-probability Eq.(\ref{LGqpI}) is non-negative, so the LG inequality holds. However, the general point here is that there is a clear logical relation between NSIT conditions and LG inequalities only if all possible NSIT conditions are satisfied, and if only some are, then the two types of conditions are no longer simply related.

We may also phrase this all in terms of the characterizations of MR for dichotomic variables outlined in Appendix A. For many-valued variables, a natural definition of strong MR at two times is to require that a suitably large set of NSIT conditions are satisfied (large enough to ensure that all interference terms vanish). Similarly weak MR at two times is the requirement that the full set of $N^2$ two-time LG inequalities hold. Clearly the former implies the latter but for many-valued variables there are then a variety of intermediate possibilities which are not simply related to each other.

The relationship between NSIT conditions and interference terms becomes more complicated for $N>3$, since it is not sufficient
to work with dichotomic variables of the form $\hat Q(n) = E_n - {\bar E}_n $. For example, for $N=4$,  there are eighteen interference terms in the decoherence functional. There are four choices of $Q(n)$ in the form defined above which means that there are twelve NSIT conditions involving $p_{12}^Q (s_1, n_2)$ which, unlike the $N=3$ case, is not enough to ensure that all interference terms are zero. What is required is a more general dichotomic variable, for example, of the form $Q=P_1 + P_2 - P_3 - P_4$.
By including two variables of this type we can bring the number of NSIT conditions up to eighteen which is enough to kill all the interference terms.



\subsection{Three-Time NSIT Conditions}

We now briefly consider three-time NSIT conditions  for the many-valued case, the generalizations of the relations, Eqs.(\ref{NSIT23})-(\ref{NSIT123b}), used in the dichotomic case. We saw in the two-time case in Section 5(B) that, for the $N=3$ case, it is possible to write down a set of NSIT conditions involving three dichotomic variables $Q$, $R$, $S$, which ensures that all interference terms are zero. Hence by analogy, we anticipate that in the three-time case for $N=3$, a complete set of NSIT conditions consists of Eqs.(\ref{NSIT23})-(\ref{NSIT123b}) in which $Q$, $R$ or $S$ is measured at the first pair of times (with a measurement $E_{n_3}$ at the final time). We will not give any more details here. For $N>3$ a more judicious choice of dichotomic variables may be required as we saw for the $N=4$ case for two times.

\section{Violations of the L\"uders Bound}

In quantum mechanics, the LG inequalities Eqs.(\ref{LG1})-(\ref{LG4}) with correlation functions given by Eq.(\ref{corr2}) have a maximum violation of $- \frac{1}{2} $ on the right-hand side. This follows from the inequality
\beq
\biggl< \left( s_1 \hat Q_1 + s_2 \hat Q_2 + s_3 \hat Q_3 \right)^2 \biggr> \ge 0,
\eeq
which is readily seen to imply that
\beq
1 + s_1 s_2 C_{12} + s_1 s_3 C_{13} + s_2 s_3 C_{23} \ge -\frac{1}{2}.
\label{Luders}
\eeq
This relation has the same mathematical form as the Tsirelson bound for the correlators in Bell-type experiments \cite{Tsi} and in the LG context this is often known  as the L\"uders bound \cite{Dak,deg,EmaExp,PQS,KQP}. 
However, unlike the Tsirelson bound for measurements on entangled pairs which represents the maximal violation permitted by quantum mechanics, 
the L\"uders bound can in fact be violated under certain circumstances. 
These violations can in principle go right up to the algebraic maximum of $-2$ on the right-hand side, accomplished for example when $C_{12} = C_{23} = C_{13} = -1 $, an outright logical paradox from the classical point of view (rather than just the statistical paradox implied by standard LG violations).

The L\"uders bound violation is possible for systems with $N \ge 3$ in which the correlator is measured in a different way.
In the usual method, one measures the dichotomic variable
\beq
\hat Q = \sum_{n} \epsilon (n) E_n,
\eeq
for some coefficients $\epsilon (n) = \pm 1 $ using a projector $P_s$ onto $\hat Q$. This is often referred to as 
a L\"uders measurement \cite{Lud}. We will refer to the resulting correlation function Eq.(\ref{corr2}) as the L\"uders correlator, $C^L_{12}$, and it has an equivalent expression in terms of the quasi-probability,
\beq 
C^L_{12} = \sum_{n_1, n_2} \epsilon (n_1) \epsilon (n_2) \ q (n_1, n_2).
\eeq
However, there is a macrorealistically equivalent method, which is to determine the two-time sequential measurement probability $p_{12} (n_1, n_2) $ using von Neumann (vN) measurements (sometimes also called ``degeneracy-breaking''), modeled by the $E_n$, and related to the L\"uders measurements by Eq.(\ref{csn}). We then construct the von Neumann correlator
\beq
C^{vN}_{12} = \sum_{n_1, n_2} \epsilon (n_1) \epsilon (n_2) \ p_{12} (n_1, n_2).
\eeq
Unlike the L\"uders correlator, LG inequalities constructed from $C^{vN}_{12}$ need not satisfy the L\"uders bound, Eq.(\ref{Luders}). 

However, we would argue, using the understanding of earlier sections, that such a violation is different in character to the usual three-time LG violations.
Using Eq.(\ref{qpN}), it is readily seen that
\beq
C^{vN}_{12} = C^L_{12} -   \sum_{n_1 \ne n_1' } \sum_{n_2} \epsilon (n_1) \epsilon (n_2)  \ I_{n_1 n_1'} (n_2).
\label{vNI}
\eeq
We thus see that the difference between the von Neumann and L\"uders correlators depends on the two-time interference terms, i.e. on the degree to which the various two-time NSIT conditions are  violated. This difference vanishes in the dichotomic case, as is readily shown.

The significance of this is as follows. As argued in Section 2, violations of the three-time LG inequalities with the usual L\"uders correlators arise entirely due to interference terms present at three times but not present at two times. Violation of the L\"uders bound using the von Neumann correlators therefore arises due to a combination of the usual three-time interferences plus some additional two-time interference terms for each time pair -- it is not due to any new kind of three-time interference term. 

This effect can then be regarded in two different ways. One attitude would be to say that since this new effect (compared to the more usual LG violations)  comes solely from two-time interference, in a systematic exploration of various MR conditions at two and three times, it might be more natural to identify these effects using two-time NSIT conditions, not the three-time inequalities Eq.(\ref{Luders}). For example, one could first explore MR conditions for measurements on all pairs of times and look for parameter ranges in which such conditions are violated or satisfied. On proceeding to the three-time case, it would then be natural to restrict only to those parameter ranges for which all two-time MR conditions hold, in order to clearly distinguish between MR conditions at two and three times. The interferences producing the L\"uders bound violation
would disappear if all two-time NSIT conditions are enforced, but a three-time LG violation up to the L\"uders bound is still possible. 


The second attitude would be to note that a L\"uders violation requires violation of both
a three-time LG inequality and a two-time NSIT condition. Hence it represents a certain economy since tests the violation of two conditions in a single experiment. In the language of Appendix A, the two conditions are strong MR at two times (which for the many-valued case we take to mean that all two-time NSIT conditions hold) and weak MR at three times.





L\"uders bounds violations are also possible with two-time LG conditions and similar comments apply in terms of the MR conditions affected. For example,
the two-time LG inequality,
\beq
1 + \langle Q_1 \rangle + \langle Q_2 \rangle + C_{12} \ge 0,
\label{LGQ++}
\eeq
and has L\"uders bound $ - \frac{1}{2}$ on the right-hand side if the correlator is the usual L\"uders one. This LG inequality can be violated if certain two-time interference terms are sufficiently large, as we have seen, for example, in Eq.(\ref{LGqpI}).
From a macrorealistic point of view
this inequality still holds if the correlator is measured using degeneracy breaking measurements. 
The correlator is then taken to be the von Neumann one, $C^{vN}_{12}$ and a violation of the two-time L\"uders bound is then possible, due to the presence of additional two-time interference terms. Hence here a L\"uders violation tests the presence of a sum of two-time interference terms see in Eq.(\ref{vNI}).

However, as stressed already, there are typically many interference terms for systems with $N \ge 3$ which can all be constrained to various degrees by various LG inequalities and NSIT conditions.
A L\"uders bound violation in this case signals a violation of a combination of certain two-time LG inequalities and certain two-time NSIT conditions.

To see this in more detail, consider the $N=3$ case discussed in detail in Section 5. We have three dichotomic variables $Q,R,S$ and we note that the LG inequality Eq.(\ref{LGQ++}) is proportional (up to a factor of $1/4$) to the quasi-probability $q(+,+)$ in Eq.(\ref{LGqpI}), which we write out explicitly:
\beq
q(+,+) = p_{12}^Q (+,+) + I_{AB}(A) + I_{AC}(A)
\eeq
The see the extra interference arising in a L\"uders violation, we compute Eq.(\ref{vNI}). We have $\epsilon (A) = 1$ and $\epsilon(B) = \epsilon (C) = -1 $ and noting that $\sum_{n_2} I_{n_1 n_1'} (n_2) = 0$, we readily find
\beq
C_{12}^{vN} = C_{12}^L - 4 I_{BC} (A)
\eeq
These relations together imply that
\beq
1 + \langle Q_1 \rangle + \langle Q_2 \rangle + C_{12}^{vN} = 4 \left( 
p_{12}^Q (+,+) + I_{AB}(A) + I_{AC}(A) - I_{BC} (A) \right)
\label{LGvn}
\eeq
and it is the presence of the term $I_{BC}(A)$ which makes a L\"uders violation possible.
The LG and NSIT conditions described in Section 5(B) will imply restrictions on all six of the interference terms in the $N=3$ case.
In particular, it is not hard to see that non-trivial values of $I_{BC}(A)$ can be identified with NSIT or conventional LG violations without having to appeal to a L\"uders violation.

Eq.(\ref{LGvn}) has another interesting feature which is that it may be negative even when the operators $\hat Q_1 $ and $\hat Q_2$ commute. The NSIT conditions Eq.(\ref{conB}) are satisfied under those conditions and so the interference terms $I_{AB}(A)$ and $I_{AC}(A)$ are zero, but since $I_{BC}(A) \ne 0$ Eq.(\ref{LGvn}) indicates that a LG violation is still possible.
This conclusion is in agreement with the earlier work Ref. \cite{KQP} in which this phenomenon was examined more generally (and also investigated the conceptual relevance of L\"uders violations, as we do here).

To be clear, none of the above remarks undermine the significance of L\"uders bound violations, which are truly striking non-classical effects, even more so than conventional LG violations. Here, we have argued here that the presence of the underlying interference terms producing them can be detected by less striking means and also identified the particular type of MR conditions that L\"uders violations test.






\section{Summary and Conclusion}

We have shown how to extend the standard conditions for macrorealism for dichotomic variables, namely two and three-time LG inequalities and NSIT conditions, to situations described by $N$-valued variables. We have in addition explored the various new features of these conditions and their relationships that do not arise in the dichotomic case.

To prepare the ground, we carried out a detailed quantum-mechanical analysis of the dichotomic case in Section 2. This highlighted the fact that conditions for MR act as constraints on the degree of interference.

In Section 3 we considered conditions for MR using LG inequalities at two times. We established a complete set of two-time LG inequalities, Eq.(\ref{LGN2Q}), in terms of the dichotomic variables $Q(n)$, which are necessary and sufficient conditions for the existence of a pairwise joint probability $p(n_1,n_2)$. The $N$ variables $Q(n)$ are however not a minimal set and we exhibited a minimal set for the case $N=3$ involving just two dichotomic variables.

MR conditions using LG inequalities at three or more times were considered in Section 4. We proved a generalization of Fine's theorem, i.e. established the necessary and sufficient conditions under which a set of three pairwise probabilities of the form $p(n_i,n_j)$ could be matched to an underlying joint probability. The conditions in question turned out to be a natural generalization of the familiar three-time LG inequalities for the dichotomic case.
We then generalized this treatment to four or more times.
We noted that Fine's ansatz for the $N$-valued variable case readily extends to the case of four or more times (as studied in the dichotomic case in Ref.\cite{HaMa}), which indicates that, like the dichotomic case, MR conditions involving LG inequalities boil down to the three-time case

In Section 5, we considered the stronger MR conditions characterized by NSIT conditions for the $N \ge 3$ case and elucidated their connection to the vanishing of certain interference terms.
We noted that  the $N  \ge 3$ case has a much richer set of NSIT conditions than the dichotomic case. In particular, Eq.(\ref{NSITn}) is not the only type of NSIT condition and many new conditions can be generated by considering measurements of different choices of dichotomic variables $Q(n)$ at the first time.
We derived the conditions in detail in the $N=3$ case and applied this understanding to some recent experiments, in which two-time LG inequality violations were observed, even though the NSIT conditions Eq.(\ref{NSITn}) were satisfied. This illustrates an important general feature: the
set of NSIT conditions and LG inequalities for many-valued variables do not have the simple hierarchical relationship enjoyed by the dichotomic case (summarized in Appendix A).

In Section 6, we used the understanding gained earlier to examine violations of the L\"uders bound for three-time LG inequalities. We noted that these entail violations of two distinct MR conditions -- a two-time NSIT condition and a three-time conventional LG violation. The effect would therefore be absent in an approach which requires all two-time MR conditions to be satisfied before proceeding to three times, but could also be regarded as an economical way of identifying two different MR condition violations in a single experiment. Similar observations were also made for the case of two-time L\"uders violations.



The present work suggests a number of new possibilities for experimental tests of macrorealism. The main one would be to test a complete set of MR conditions for a system with $N \ge 3$, as was recently carried out for the dichotomic case \cite{Maj}. This would involve making measurements which are able to check all of the two-time and three-time LG inequalities, Eqs.(\ref{LGN2Q}), (\ref{LGN3Q}), thereby checking for violations of MR at both two times and three times. A limited number of LG tests involving three-level systems have been carried out  but none test a complete set of two and three time MR conditions (although the one recent experiment discussed in Section 3 which tested a complete set of two-time conditions \cite{EmINRM} is an important step forwards here).
The proposed new experiments could readily be carried out by simple extensions of existing approaches. 
We also note that the recent proposal to test MR in the context of the harmonic oscillator, involving a dichotomic variable equal to the sign of the position operator \cite{Bose}, could readily be extended to many-valued variables by partitioning the position into more than two values. 
Further tests of the L\"uders bound are also of interest.


\section{Acknowledgements}

We are grateful to Sougato Bose, Clive Emary, Dipankar Home, George Knee, Johannes Kofler, Raymond Laflamme, Shayan Majidy, Owen Maroney, Alok Pan, Stephen Parrott and James Yearsley for many useful discussions and email exchanges about the Leggett-Garg inequalities over a long period of time. We also thank Shayan Majidy for a critical reading of the manuscript. In addition, we thank two anonymous referees for helpful comments.

\appendix

\section{Detailed Characterization of Different Types of Macrorealism}


So far we have discussed conditions for MR based on LG inequalities and NSIT conditions, which, as indicated in the Introduction, are different types of conditions for MR. For clarity, we now make these different definitions more precise, following Refs.\cite{HalQ,HalLG4}.
As stated already, Eq.(\ref{MRdef}), macrorealism is the logical conjunction of NIM, MRps and induction. 
The last requirement, induction, is rarely contested. The first two can be interpreted in a number of different ways, so elaboration is required.

First of all, MRps comes in three different types \cite{MaTi}, but the LG framework tests only one of them, which are essentially hidden variable theories of the GRW type \cite{GRW}. The other two types are hidden variable theories in which the wave function itself is included in the specification of the ontic state, as is the case, for example, in de Broglie-Bohm theory, and models of this type are much harder to rule out experimentally,
Secondly, the NIM requirement may also be interpreted in a number of different ways depending on how many quantities are determined in each individual experiment. We illustrate this for MR tests at both two and three times.

At two times, there are two natural ways to proceed. One is to measure the two-time probability $p_{12} (s_1,s_2)$ using sequential measurements in a single experiment and then require that it satisfies the NSIT condition Eq.(\ref{NSIT}), which we denote $\NSIT_{(1)2}$ \cite{KoBr}.
We refer to the version of NIM involved in such a test as {\it sequential} NIM, denoted $\NIM_{seq}$, and
refer to the definition of MR tested in this way as strong MR:
\beq
\MR_{strong} = \NSIT_{(1)2} \wedge \Ind.
\label{MRs}
\eeq
Note that the NSIT condition embraces both NIM and MRps in this approach, since it both shows that the first of the two measurements does not disturb the second and supplies the joint probability for the pair of measurements.

The other way is to measure $\langle Q_1 \rangle$, $\langle Q_2 \rangle$ and $C_{12}$ in three different experiments, requiring non-invasiveness in each individual experiment. The only non-trivial issue in terms of invasiveness is the measurement of $C_{12}$ which is carried out using standard methods in LG experiments, as described in Section 2(A). Crucially, here we do not demand that that the measurements in different experiments would still be non-invasive if combined. That is, non-invasiveness is maintained only in each separate piece of the data set, but not the whole.
(This is a direct analogy to what is done in Bell experiments). We therefore refer to this version of NIM as {\it piecewise}, and denote it $\NIM_{pw}$.
Since $\NIM_{pw}$ is clearly weaker than $\NIM_{seq}$, the corresponding version of MR is much weaker, and we denote it,
\beq
\MR_{weak} = \NIM_{pw} \wedge \LG_{12} \wedge \Ind.
\label{MRw}
\eeq

At three times one may proceed similarly. One may use $\NIM_{seq}$ and measure the three-time probability $p_{123} (s_1,s_2,s_3)$ directly using three sequential measurements in a single experiment
and then require that it satisfies a set of NSIT conditions \cite{KoBr,Cle}. A suitable set are the conditions,
\bea
\sum_{s_2} p_{23} (s_2,s_3) &=& p_3( s_3),
\label{NSIT23}
\\
\sum_{s_1} p_{123} (s_1,s_2,s_3) &=& p_{23} (s_2,s_3),
\label{NSIT123a}
\\
\sum_{s_2} p_{123} (s_1,s_2,s_3) &=& p_{13} (s_1,s_3), 
\label{NSIT123b}
\eea
which are denoted $\NSIT_{(2)3}$, $\NSIT_{(1)23}$ and $\NSIT_{1(2)3}$ respectively.
The corresponding definition of strong MR is
\beq
\MR_{strong} = \NSIT_{(2)3} \wedge \NSIT_{(1)23} \wedge \NSIT_{1(2)3} \wedge \Ind.
\eeq
Alternatively, one can work with $\NIM_{pw}$ in which the results of a number of different experiments are combined. In particular, six experiments are carried out to determine the three $\langle Q_i \rangle$ and three $C_{ij}$, so no more than two measurements are made in each individual experiment. Weak MR is then defined using a combination of two and three time LG inequalities:
\beq
\MR_{weak} = \NIM_{pw} \wedge \LG_{12} \wedge  \LG_{23} \wedge \ LG_{13}  \wedge \LG_{123} \wedge \Ind.
\eeq
An intermediate possibility is to use $\NIM_{seq}$ for the two-time measurements, but use $\NIM_{pw}$ in assembling the two-time probabilities into a three time probability, leading to the definition of MR:
\beq
\MR_{int} = \NSIT_{(1)2} \wedge \NSIT_{(1)3} \wedge \NSIT_{(2)3} \wedge \LG_{123} \wedge \Ind
\eeq
Most LG experiments to date appear to be testing either $\MR_{weak}$ or $\MR_{int}$ although this is not necessarily made clear in many previous works.

There is a clear hiearchy in these conditions, namely
\beq
\MR_{strong} \implies \MR_{int} \implies \MR_{weak}
\eeq
which follows because the NSIT conditions imply that LG inequalities must hold but the converse is not true. However, this hierarchy is specifically for the case of measurements of a single dichotomic variable. 
For the case of many-valued variables we consider in this paper there is a richer variety of NSIT conditions and LG inequalities and consequently a richer variety of intermediate MR conditions which do not have a straightforward hierarchical relationship.

\section{Some results from the decoherent histories approach to quantum mechanics}

Here  we briefly outline some properties of the decoherence functional and its relation to the quasi-probability and sequential measurement formula. These results are standard mathematical ones from the decoherent histories approach to quantum theory \cite{HalQ,GH2,Gri,Omn1,HalQIP} although this is not a decoherent histories analysis. 

Histories consisting of  measurements at $n$ times are represented by ``class operators'',
\beq
C_{\alpha} = P_{s_n} (t_n) \dots P_{s_2} (t_2) P_{s_1} (t_1),
\eeq
where $\alpha$ denotes the string $(s_1,s_2, \cdots s_n)$. They sum to the identity. Each class operator has negation defined by
\beq 
\overline{C}_{\alpha} = 1 - C_{\alpha} = \sum_{\alpha' \ne \alpha} C_{\alpha'},
\eeq
which therefore represents all the histories not corresponding to the string $(s_1,s_2, \cdots s_n)$.
The probability for a sequence of measurements at $n$ times is
\beq
p(\alpha) = {\rm Tr} \left( C_{\alpha} \rho C_{\alpha}^\dag\right),
\eeq
and the associated quasi-probability is
\beq
q(\alpha) ={\rm Re} {\rm Tr} \left( C_{\alpha} \rho  \right).
\eeq
We also introduce the decoherence functional
\beq
D(\alpha, \alpha') = {\rm Tr} \left( C_{\alpha} \rho C_{\alpha'}^\dag\right),
\eeq
describing interference between the history $C_{\alpha}$ and history $C_{\alpha'}$.
Simple algebra then shows that
\beq
q(\alpha) = p(\alpha) + {\rm Re} D(\alpha, \bar \alpha),
\label{pqd}
\eeq 
where 
\beq
D(\alpha, \bar \alpha) = {\rm Tr} \left(C_{\alpha}\rho  \overline{C}_{\alpha}^\dag \right)
\eeq
is the decoherence functional describing the interference between history $C_{\alpha}$ and its negation $ \overline{C}_{\alpha} $.
Eq.(\ref{pqd}) may also be written,
\beq
q(\alpha) = p(\alpha) + \sum_{{\alpha'} \atop {\alpha' \ne \alpha}} {\rm Re} D(\alpha, \alpha').
\label{pqd2}
\eeq

\section{Leggett-Garg Inequalities at Three Times for the $N=3$ Case}

We give here the explicit form for the three-time LG inequalities Eq.(\ref{LGN3Q}) in the $N=3$ case, in terms of the three dichotomic variables $Q$, $R$ and $S$, satisfying $Q+R+S=-1$. They are:
\begin{align}
1+\expval{Q_1 Q_2}+\expval{Q_2 Q_3}+\expval{Q_1 Q_3}&\geq 0,\\
1+\expval{R_1 Q_2}+\expval{Q_2 Q_3}+\expval{R_1 Q_3}&\geq 0,\\
1+\expval{Q_1 R_2}+\expval{R_2 Q_3}+\expval{Q_1 Q_3}&\geq 0,\\
1+\expval{R_1 R_2}+\expval{R_2 Q_3}+\expval{R_1 Q_3}&\geq 0,\\
1+\expval{Q_1 Q_2}+\expval{Q_2 R_3}+\expval{Q_1 R_3}&\geq 0,\\
1+\expval{R_1 Q_2}+\expval{Q_2 R_3}+\expval{R_1 R_3}&\geq 0,\\
1+\expval{Q_1 R_2}+\expval{R_2 R_3}+\expval{Q_1 R_3}&\geq 0,\\
1+\expval{R_1 R_2}+\expval{R_2 R_3}+\expval{R_1 R_3}&\geq 0,\\
1+\expval{S_1 Q_2}+\expval{Q_2 Q_3}+\expval{S_1 Q_3}&\geq 0,\\
1+\expval{S_1 R_2}+\expval{R_2 Q_3}+\expval{S_1 Q_3}&\geq 0,\\
1+\expval{S_1 Q_2}+\expval{Q_2 R_3}+\expval{S_1 R_3}&\geq 0,\\
1+\expval{S_1 R_2}+\expval{R_2 R_3}+\expval{S_1 R_3}&\geq 0,\\
1+\expval{Q_1 S_2}+\expval{S_2 Q_3}+\expval{Q_1 Q_3}&\geq 0,\\
1+\expval{R_1 S_2}+\expval{S_2 Q_3}+\expval{R_1 Q_3}&\geq 0,\\
1+\expval{Q_1 S_2}+\expval{S_2 R_3}+\expval{Q_1 R_3}&\geq 0,\\
1+\expval{R_1 S_2}+\expval{S_2 R_3}+\expval{R_1 R_3}&\geq 0,\\
1+\expval{Q_1 Q_2}+\expval{Q_2 S_3}+\expval{Q_1 S_3}&\geq 0,\\
1+\expval{R_1 Q_2}+\expval{Q_2 S_3}+\expval{R_1 S_3}&\geq 0,\\
1+\expval{Q_1 R_2}+\expval{R_2 S_3}+\expval{Q_1 S_3}&\geq 0,\\
1+\expval{R_1 R_2}+\expval{R_2 S_3}+\expval{R_1 S_3}&\geq 0,\\
1+\expval{S_1 S_2}+\expval{S_2 R_3}+\expval{S_1 R_3}&\geq 0,\\
1+\expval{S_1 S_2}+\expval{S_2 Q_3}+\expval{S_1 Q_3}&\geq 0,\\
1+\expval{Q_1 S_2}+\expval{S_2 S_3}+\expval{Q_1 S_3}&\geq 0,\\
1+\expval{R_1 S_2}+\expval{S_2 S_3}+\expval{R_1 S_3}&\geq 0,\\
1+\expval{S_1 Q_2}+\expval{Q_2 S_3}+\expval{S_1 S_3}&\geq 0,\\
1+\expval{S_1 R_2}+\expval{R_2 S_3}+\expval{S_1 S_3}&\geq 0,\\
1+\expval{S_1 S_2}+\expval{S_2 S_3}+\expval{S_1 S_3}&\geq 0.
\end{align}
Due to the non-minimal nature of the set $Q, R, S$, we in fact do not need to measure all 
twenty-seven correlators, since all averages and correlators involving $S$ may be expressed in terms of $Q$ and $R$. So we have for example,
\beq
\expval{S_1 Q_2} = - \expval{Q_2} - \expval{Q_1 Q_2} - \expval{R_1 Q_2},
\eeq
and also
\bea
\expval{S_1 S_2}=1 &+& \expval{Q_1}+\expval{Q_2}+\expval{R_1}+\expval{R_2}
\nonumber \\
&+& \expval{Q_1 Q_2}+\expval{Q_1 R_2}+\expval{R_1 Q_2}+\expval{R_1 R_2}.
\eea
All other cases have this general form and we will not write them out here. The set of quantities to be measured then consists of the twelve correlators of the form $\expval{Q_i R_j}$, $\expval{Q_i Q_j}$, $\expval{R_i R_j}$ (the last two with $i<j$), along with the six averages $\expval{Q_i}$ and $\expval{R_i}$, for $i,j =1,2,3$.

\bibliography{apssamp}

\begin{thebibliography}{10}


\bibitem{LeGa} A.J.Leggett and A.Garg,
Phys. Rev. Lett. 54, 857 (1985).


\bibitem{L1} A. J. Leggett,
Found. Phys. 18, 939 (1988); 
Rep. Prog. Phys. 71, 022001 (2008).


\bibitem{ELN}  C. Emary, N. Lambert and F. Nori, Rep. Prog. Phys. 77, 016001 (2014)


\bibitem{Bell} J.S.Bell, Physics (N.Y.) 1, 195 (1964), reprinted, along with most
of Bell's other key papers, in J.S.Bell, {\it Speakable and Unspeakable in Quantum
Mechanics} (Cambridge University Press, Cambridge, 1987).

\bibitem{CHSH} J.F.Clauser, M.A.Horne, A.Shimony and R.A.Holt, Phys.Rev.Lett. 23, 1306 (1982);
J.F.Clauser and A.Shimony, Rep. Prog. Phys. 41, 1881 (1978).






\bibitem{MaTi} O.J.E Maroney and C.G Timpson, arXiv:1412.6139 (2014).


\bibitem{Fine} A.Fine, J.Math.Phys. 23, 1306 (1982); Phys.Rev.Lett. 48, 291 (1982).
Some of Fine's work has overlaps with the earlier work: N. N. Vorobev, Theory Probab. Appl., 7(2), 147–163 (1959).



\bibitem{Bus} P.Busch, in {\it Non-locality and Modality} edited by. T. Placek, J. Butterfield, Springer-Verlag, NATO Science Series II. Mathematics, Physics and Chemistry
64, 175 (2002). (Also available as quant-ph/0110023).

\bibitem{SuZa} P.Suppes and M.Zanotti, Synthese 48, 191 (1981).

\bibitem{Pit} I.Pitowski, {\it Quantum Probability -- Quantum Logic}, Lecture Notes in Physics 321 (Springer-Verlag, Berlin, 1989).

\bibitem{GaMer} A.Garg and N.D.Mermin, Found.Phys. 14, 1 (1984).

\bibitem{ZuBr} M.Zukowski and C.Brukner, Phys. Rev. Lett. 88, 210410 (2002).

\bibitem{JHFine} J. J. Halliwell, Phys. Lett. A 378, 2945 (2014).



\bibitem{AbBr} S.Abramsky and A.Brandenburger, New Journal of Physics, 13, 113035
(2011).



\bibitem{HalQ} J.J.Halliwell, Phys. Rev. A 93, 022123 (2016).

\bibitem{HalLG4} J.J.Halliwell, Phys. Rev. A 96, 012121 (2017). A concise and updated version of this work is the e-print arXiv:1811.10408.


\bibitem{Maj} 
S. Majidy,  H. Katiyar, G. Anikeeva,  J. J. Halliwell and R. Laflamme, Phys. Rev. A 100, 042325 (2019).


\bibitem{MajThe} S. Majidy, {\it Violation of an augmented set of Leggett-Garg inequalities and the implementation of a continuous in time velocity measurement}, Master's Thesis (University of Waterloo, 2019).



\bibitem{HaMa} 
J.J.Halliwell and C.Mawby, Phys. Rev. A 100, 042103 (2019).

\bibitem{HalNIM} J. J. Halliwell, Phys. Rev. A 99, 022119 (2019).


\bibitem{PQS} A. K. Pan, Md. Qutubuddin and S. Kumari, Phys. Rev. A 98, 062115 (2018).

\bibitem{poly} M. Ara\'ujo, M. T\'ulio Quintino, C. Budroni, M. Terra Cunha, and Ad\'an Cabello, 	Phys. Rev. A 88, 022118 (2013).


\bibitem{KoBr} J. Kofler and C. Brukner,
Phys. Rev. A 87, 052115 (2013).

\bibitem{Cle} L.Clemente and J.Kofler, Phys. Rev. A 91, 062103 (2015); Phys. Rev. Lett. 116, 150401 (2016).



\bibitem{WLG} D. Saha, S. Mal, P. K. Panigrahi and D. Home, Phys. Rev. A, 91, 032117 (2015);
S. Kumari and A. K. Pan, 
Phys. Rev. A 96, 042107 (2017); S. Kumari and A. K. Pan, 
EPL, 118, 50002 (2017).

\bibitem{KuPa} S. Kumari and A. K. Pan, arXiv:1912.10977.  

\bibitem{Bose}
S. Bose, D. Home and S. Mal, Phys. Rev. Lett. 120, 210402 (2018)


\bibitem{Dak} B. Dakic, T. Paterek, and C. Brukner, New J. Phys. 16, 5023028 (2014).

\bibitem{deg} C. Budroni and C. Emary, Phys. Rev. Lett. 113, 050401 (2014).


\bibitem{EmaExp} K. Wang, C. Emary, X. Zhan, Z. Bian, J. Li and  P. Xue,
Opt. Exp. 25, 31462 (2017).


\bibitem{KQP} 
A.Kumari, Md. Qutubuddin and A. K. Pan, Phys. Rev. A 98, 042135 (2018).

\bibitem{Tsi} B. S. Tsirelson, 
Lett. Math. Phys. 4, 93 (1980).



\bibitem{GoPa} S.Goldstein and D.N.Page, Phys.Rev.Lett 74, 3715 (1995).


\bibitem{HaYe} J.J.Halliwell and J.M.Yearsley, Phys.Rev. A87, 022114 (2013).

\bibitem{Kly} D.N.Klyshko, Phys. Lett. A218, 119 (1996).



\bibitem{Fri} T. Fritz,
New J. Phys. 12, 083055 (2010).


\bibitem{GH2} M.Gell-Mann and J.B.Hartle, Phys.Rev. { D47},
3345 (1993).


\bibitem{Gri} R.B.Griffiths, J.Stat.Phys. { 36}, 219 (1984);
Phys.Rev.Lett. { 70}, 2201 (1993); Phys.Rev. { A54}, 2759
(1996); { A57}, 1604 (1998).

\bibitem{Omn1} R. Omn\`es, J.Stat.Phys. { 53}, 893 (1988).
{ 53}, 933 (1988); { 53}, 957 (1988); { 57}, 357 (1989);
Ann.Phys. { 201}, 354 (1990); Rev.Mod.Phys. { 64}, 339
(1992).

\bibitem{HalQIP} J. J. Halliwell, Quant. Info. Proc. 8, 479 (2009).


\bibitem{Wit} 
C.-M. Li, N. Lambert, Y.-N. Chen, G.-Y. Chen, and F. Nori,
Sci. Rep. 2, 885 (2012);
G. Schild and C.Emary, Phys. Rev. A 92, 032101 (2015); 
K. Wang, G. C. Knee, X.  Zhan, Z. Bian, J.  Li and P. Xue,
Phys. Rev. A 95, 032122 (2017).

\bibitem{HalCTVM} J. J. Halliwell,  Phys. Rev. A 94, 052114 (2016).

\bibitem{Ema} C.Emary, Phys. Rev. A 96, 042102 (2017).



\bibitem{EmINRM}  K. Wang, C.  Emary, M. Xu, X. Zhan, Z. Bian, L. Xiao and  P.Xue,
Phys. Rev. A 97, 020101 (2018).

\bibitem{3box} R. E. George, L. M. Robledo, O. J. E. Maroney, M. S. Blok,
H. Bernien, M. L. Markham, D. J. Twitchen, J. J. L. Morton,
G. A. D. Briggs, and R. Hanson, 
Proc. Natl. Acad. Sci. USA 110, 3777 (2013).




\bibitem{HalDS} J. J. Halliwell,  A. Bhatnagar, E. Ireland, H. Nadeem and V. Wimalaweera, {\it The Double Slit Experiment as a Leggett-Garg Test} (in preparation).

\bibitem{Lud}   G. L\"uders, Ann. Phys. (Leipzig) 8, 322-328 (1951). A translation may be found at
arXiv:quant-ph/0403007.





















\bibitem{GRW} G. C. Ghirardi, A. Rimini, and T. Weber.
Phys. Rev. D 34, 470 (1986);
P. Pearle,
Phys. Rev. A 39, 2277 (1989);
R. Penrose, 
General Relativity and Gravitation,
28, 581 (1996);
A. Bassi, K. Lochan, S. Satin, T. P. Singh and H. Ulbricht, 
Rev Mod Phys 85, 471–527 (2013).





\end{thebibliography}

\end{document}